# A spin quantum bit architecture with coupled donors and quantum dots in silicon

T. Schenkel[1], C. C. Lo[1], C. D. Weis[1], J. Bokor[2], A. M. Tyryshkin[3], and S. A. Lyon[3]

[1]Ion Beam Technology Group, Lawrence Berkeley National Laboratory,
Berkeley, CA 94720, USA

[2]Department of Electrical Engineering and Computer Sciences, University of California,
Berkeley, CA 94720, USA

[3]Electrical Engineering Department, Princeton University, Princeton, NJ, USA

Contact-email: T_Schenkel@LBL.gov

(October 10, 2011)

Spins of donor electrons and nuclei in silicon are promising quantum bit (qubit) candidates which combine long coherence times with the fabrication finesse of the silicon nanotechnology industry. We outline a potentially scalable spin qubit architecture where donor nuclear and electron spins are coupled to spins of electrons in quantum dots and discuss requirements for donor placement aligned to quantum dots by single ion implantation.

## 1. INTRODUCTION

Electron and nuclear spins of donors in silicon have long been recognized as promising qubit candidates [1]. In isotopically purified $^{28}$Si they exhibit long coherence times [2, 3] and their integration can benefit from the great fabrication finesse of silicon nanotechnology. Several prominent proposals for scalable quantum computer architectures with donor spin qubits have emerged [1, 4-7]. In the original Kane proposal, quantum information is stored in the nuclear spin of phosphorus atoms. Electrostatic gates facilitate transfer of quantum information form nuclear to electron spins and between electron spins, by modulation of the contact hyperfine interaction (A-gates), and the exchange coupling (J-gates), respectively. Recently, reliable detection of single electron spins [8] and the control of single electron and nuclear spins [9] states was reported for donors in silicon. Similar advances have also been reported for NV$^-$ centers in diamond [10, 11] where there are many materials and integration challenges complementary to those for donors in silicon [12, 13].

Following Di Vincenzo's widely used criteria for the development of a large scale quantum computer [14], elements of quantum memory, quantum logic and efficient quantum communication channels have to be integrated. While single donor electron and nuclear spin readout and control have been demonstrated, the next difficult challenges are to master spin qubit coupling so that two and multi-qubit logic operations can be implemented. In early donor qubit proposals, coupling was envisioned along 1D chains of nearest neighbor coupled qubits. This ought to suffice for quantum logic demonstrations with several qubits even with limited coupling control [15] but severe limitations of nearest neighbor coupling have been pointed out [4, 16]. Coherent shuttling of electrons between donors has been proposed as a path to circumvent nearest neighbor coupling challenges or to supplement nearest neighbor coupling with a longer range coupling option [4, 17]. For electron shuttling, critical questions regard spin coherence of donor electron and nuclear spins during cycles of ionization and recombination. Other potential paths for long range transport of quantum information from donor spins include concepts of a spin bus [18], virtual phonon mediated coupling [19], coupling via nano-mechanical resonators [20] and spin to photon coupling in optical cavities [21] or via high Q microwave resonators [10, 22, 23].

In parallel to single donor spin control, control of electron spins in silicon and SiGe based quantum dots has also matured rapidly [24-28]. Here, quantum information can be encoded e. g.

in the spin state of a coupled pair of electrons in a double quantum dot structure. For spin based quantum computers with donors, the nuclear spin stands out as a promising resource for quantum memory [2]. Electrons of donors and dots allow fast single qubit operation and nearest neighbor two qubit interactions through controlled exchange coupling. Cluster state quantum computer approaches offer an alternative approach e. g. with nuclear spin memory [29] that "only" require nearest neighbor interactions and reliable single qubit control and readout.

In the following we outline a quantum computer architecture where donor nuclear spins are coupled via donor electron spins to spins of electrons in quantum dots, enabling further coupling e. g. to high Q resonators for quantum communication or enabling cluster state QC implementations without need for donor ionization and coherent recombination. We then discuss elements for practical implementation of such a donor-dot architecture, i. e. fabrication of back gated quantum dots, vertically aligned to single donors in isotopically purified $^{28}$Si.

## 2. COUPLED DONOR-QUANTUM DOT SPIN QUBITS

Figure 1 shows a schematic of the envisioned spin qubit architecture with vertically coupled donors and quantum dots formed e. g. in SOI (silicon on insulator) with a thin buried oxide (box). The basic idea is to integrate the following modules and interactions:
- a single donor nuclear spin memory
- coupled to the donor electron spin through the contact hyperfine interaction for storage and retrieval of quantum information [2, 30]
- the donor electron spin coupled to the spin of an electron in a vertically aligned quantum dot, with gate controlled exchange coupling.

In effect, electron spins in quantum dots are coupled to a nuclear spin memory. Once in the quantum dot domain, a broad range of opportunities has been outlined for further coupling, from 1D chains [18, 31] to promising routes for mid and long range coupling of quantum dots [10, 23]. Readout of quantum information can be performed on either the donor or quantum dot electron spins, e. g. through spin dependent tunneling [8, 24]. Single qubit operations can be performed on the donor electron spin and on the dot electron spin with pulsed microwaves, which can be delivered locally [24] or globally by placing devices into microwave cavities [32-34]. Two qubit operations can be implemented between spins of donor and dot electrons and on electron spins of adjacent dots. Here, a large enough (i. e. >10 mG/nm) magnetic field must be applied to detune the spin resonance lines for adjacent donor-dot devices, e. g through the presence of micro-magnets [35] or local inductors. Further integration for mid and long range quantum communication through electron shuttling between dots or coupling to superconducting resonators is possible and combines all three critical architecture elements of memory, logic and communication.

An alternative approach of cluster state quantum computing can also be implemented here without the need for cycles of coherent donor ionization and recombination. Ionizing the donor protects the nuclear spin from decoherence through uncontrolled interaction with the donor electron [9, 36], but it remains unclear if coherence can be preserved in the recombination step that is necessary to retrieve the quantum information from the nuclear spin and transfer it back to the donor electron spin for further processing. Inter-dot coupling can effectively entangle donor nuclear spins and enable implementation of cluster state quantum computing [29].

While key elements of this donor – dot architecture have been tested experimentally - at least in ensemble measurements– a great many challenges remain. E. g. efficient quantum information transfer between donor electron and nuclear spins was demonstrated with ensembles of phosphorus donors in $^{28}$Si by Morton et al. [2]. Also, quantum dots with a high degree of control have recently been demonstrated in silicon and Si-SiGe hetero-structures, i. e. in materials systems that can be prepared with minimal nuclear spin background [37, 38]. Further, these nuclear spin free matrixes can be prepared on insulator layers, enabling back gating of devices in SOI (silicon on insulator, where $^{28}$SOI was demonstrated in [38]) and SGOI (Silicon Germanium on insulator, which has not yet been prepared with isotope purification) [39].

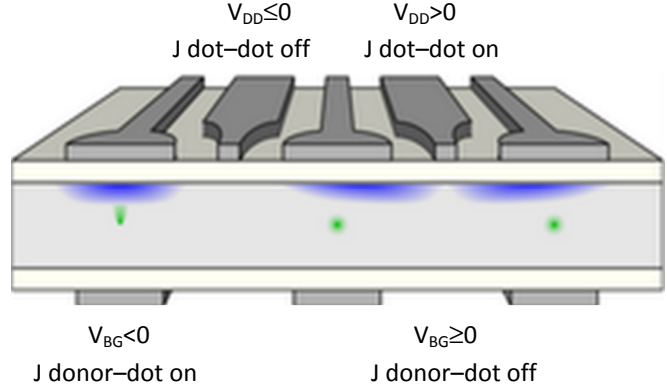

Figure 1: Schematic of a hybrid donor – quantum dot qubit architecture. Top gate defined quantum dots (blue) with single electron occupancy control are formed on a high quality dielectric aligned to donors (green) ~20 to 30 nm below. Local back gates below the box allow tuning of the donor – dot exchange coupling. Donor – donor distances are ~100 nm, large enough to avoid spurious direct donor-donor coupling, while J coupling between dots is controlled with top gates.

The six-fold valley degeneracy of the silicon conduction band leads to oscillations of the exchange coupling of adjacent donors as a function of donor – donor distance [40]. The degeneracy is partially lifted by strain. The extend of J-coupling oscillations and the effective spin-valley interplay are concerns for this scheme of donor – dot coupling and are subject of experimental and theoretical investigation. Detailed theoretical analysis is required to set bounds for defect tolerance in this architecture.

In the following we briefly discuss coherence times of donor electron and nuclear spins in $^{28}$Si and then outline device fabrication elements and requirements for single ion placement for the formation of donor – quantum dot qubit devices.

## 3. COHERENCE OF DONOR SPINS IN 28-SILICON

Preparation of a nuclear spin free matrix reduces spectral diffusion from a nuclear spin bath and leads to large increases of electron spin coherence times of donors in silicon. Spurious donor-donor coupling is reduced in dilutely doped samples [41] and $T_{2e}$ can reach values in excess of seconds [3]. Corresponding nuclear spin coherence times in highly purified $^{28}$Si are even longer, reaching ~150 s in the highest quality material [42]. Donor nuclear coherence is limited by donor electron spin relaxation, $T_{2n} \leq 2\ T_{1e}$, i. e. electron spin flips decohere the nuclear spin. Donor nuclear spin relaxation times have been found to exceed $T_{1e}$ by a factor of ~250, i. e. $T_{1n}=250\ T_{1e}$ [43]. Electron spin relaxation times exceed hundreds of seconds at temperatures below ~2 K. Further, donor electron spin relaxation times decrease rapidly with increasing magnetic fields, $T_{1e} \sim B^{-5}$ (see e. g. [44]). For electrons in quantum dots, this scaling is $T_{1e} \sim B^{-7}$ [26]. High magnetic fields lead to high decrees of spin polarization at low temperature and large Zeeman splittings enable single spin readout through spin dependent tunneling of donor electrons into quantum dots or for tunneling between coupled quantum dots [24]. Injecting spin polarized electrons from ferro-magnetic contacts across suitable tunnel barriers and trapping them in quantum dots is a potential alternative to the use of high magnetic fields to achieve electron polarization. For single spin readout, alternatives viable at low magnetic fields can be envisioned, e. g. through variants of spin dependent charge transfer into D$^-$ states[45-47], e. g. from quantum dots into donors.

Device integration of donors in a transistor paradigm, e. g. with local gate control of single and two qubit interactions, requires integration with electrodes, which can be isolated from the matrix with thin dielectrics. Imperfect interfaces increase magnetic noise and lead to much reduced coherence times. In pulsed x-band ESR experiments we have found that the electron spin coherence of $^{121}$Sb donors at 5 K is limited to just 0.3 ms for donors implanted to a mean depth of 50 nm in a $^{28}$Si epi layer [48]. The antimony fluence was $2 \times 10^{11}$ cm$^{-2}$ and samples were annealed at 980°C for 7 s to activate the donors and to repair the implant damage. When increasing the implantation depth to 150 nm, $T_{2e}$ increased to 1.5 ms. Here, the thermal oxide had an interface charge density of $\sim 1 \times 10^{11}$ cm$^{-2}$. A factor of at least ten improvement of this interface quality is possible with optimized processing. When the oxide was removed in hydrofluoric acid, the silicon surface became hydrogen passivated and $T_{2e}$ increased to 0.75 ms for the 50 nm deep donors and to 2.1 ms for the donors with a mean depth of 150 nm. The underlying physical mechanisms that limit coherence of donor electron spins at the Si-SiO$_2$ interface are not well understood. E. g. studies of the temperature dependence of $T_{2e}$ for a series of interface qualities will aid differentiation of underlying coherence limiting processes. A detailed model suggests a dominant role of dangling-bond spin relaxation and magnetic 1/f noise [49].

Very recently, these studies were extended to probe nuclear spin coherence for electrons close to the Si-SiO$_2$ interface and in preliminary results $T_{2n}$ was found to be about 40 ms at 5 K for a shallow antimony implant in a $^{28}$Si epi layer with a peak concentration of $5 \times 10^{16}$ Sb-atoms/cm$^3$ at 5 K [50]. This coherence time is over three orders of magnitude shorter then the bulk reference value. In measurements at lower temperatures and with optimized interfaces much longer $T_{2n}$ values of donors close to the Si-SiO$_2$ interface can be anticipated. Even at a few tens of ms, the nuclear spin makes an attractive quantum memory, provided that reliable quantum information transfer between the donor electron and nuclear spin is accomplished at least 10$^4$ times faster to enable application of error correction schemes [2, 29, 51].

When donor electrons are exposed to conduction electrons, e. g. from a two dimensional electron gas (2DEG) in the channel of a field-effect transistor, spin coherence times can be expected to be even shorter then in the presence of an SiO$_2$ interface alone. We have conducted systematic studies where the magnetic resonances from 2DEG and donor bound electrons where detected electrically. Electrical detection of magnetic resonance (EDMR) shows increased sensitivity compared to standard Electron Spin Resonance (ESR), which requires about 10$^9$ spins per resonance line. In our EDMR studies in X- and W-band cavities [32, 33, 52], micron scale transistors where formed in $^{28}$Si epi layers. Donor were implanted into transistor channels. Resonant current changes where observed for 2DEG electrons close to the free electron g-factor and for the hyperfine split electron spin resonance lines corresponding to the nuclear spin configuration of the isotopes used (e. g. $^{121}$Sb, nuclear spin I=5/2) (Figure 2). Resonant current changes originate from rapid exchange of spin polarization of donors to the 2DEG and the magneto-resistance of the 2DEG. Donors electrons are depolarized on resonance. Through exchange scattering with conduction electrons, donors then depolarize the 2DEG to a degree. Due to the magneto-resistance of the 2DEG, i. e. the change of channel current as a function of magnetic field, this depolarization changes the channel current when the donor electron spins are in resonance. The magnitude of the resonant current change from donors, dI/I, was about $2 \times 10^{-4}$ at W-band (94 GHz) and 5 K [52]. From studies of the microwave power dependence of the line widths of the donor and 2DEG resonances we can extract values for the 2DEG and donor electron spin coherence times. The line widths, δB, of the donors and 2DEG were 0.6 G and 1.5 G, respectively, and had Lorenzian shapes, indicating they were not limited by inhomogeneous broadening. With

$$\delta B = \frac{1}{\sqrt{3}} \frac{h}{\pi g \mu_B} \frac{1}{T_{2e}}$$

(h: Planck constant, g: Landé g-factor, $\mu_B$: Bohr magneton), we find corresponding spin coherence times of 40 ns for 2DEG and ~100 ns for donor electrons interacting with a 2DEG [53]. These short coherence times discourage 2DEG mediated coupling of donors, e. g. through an effective RKKY type interaction.

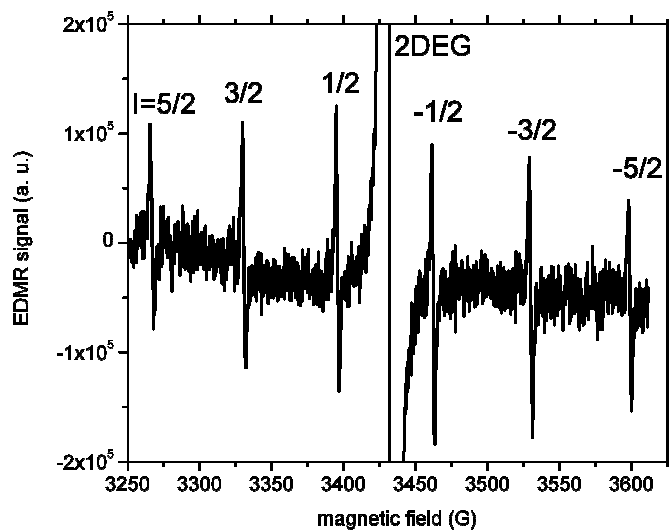

Figure 2: X-band EDMR spectrum from a 20 x 160 μm² FET where the channel had been implanted with antimony [32].

Donor spin relaxation times from the same analysis showed $T_{1e}$ of order 300 ns [53]. This value for donor electron spin relaxation in the presence of a 2DEG allows us to estimate the potential for a quantum non-demolition readout of a single nuclear spin state in an sub-100 nm scale FET [38] by EDMR [30]. If we assume that $T_{1n} \approx 250\, T_{1e}$ also for donors exposed to a 2DEG current [43], then the nuclear spin readout time is restricted to <75 μs. With resonant current changes of $2 \times 10^{-4}$, the corresponding signal levels are of order of the shot noise in nano-scale devices [30, 38, 53].

The quality of the $SiO_2$-Si interface is of critical importance, especially also for quantum dots, both for electron spin coherence and for control of single electron dot occupancy. While it can be expected that coherence limiting noise sources, such as magnetic fluctuators at the interface, will freeze out to some degree at lower temperatures, it is clear that interface quality is a critical factor for both donor and quantum dot electron spin qubit integration. Table 1 summarizes some of the constrains and relations between electron and nuclear spin coherence.

Hydrogen passivation of silicon provides the highest quality interface [57] with the longest electron spin coherence times of nearby donors [48] but technical challenges are sever to integrate it with 20 to 50 nm scale top gates and e. g. a ~10 nm scale vacuum gap.

Protection of electron spins from interface noise can be achieved by forming quantum dots in strained silicon quantum wells in Si-SiGe hetero-structures, where top gating of a high mobility silicon 2DEG >100 nm away from the surface has been reported [27, 58] including in nuclear spin free $^{28}$Si-$^{28}$Si$^{70}$Ge structures [37]. Here, donors would be implanted into the relaxed $^{28}$Si$_{0.7}$$^{70}$Ge$_{0.3}$ buffer layer below a $^{28}$Si quantum well. Compatibility of ion implantation steps with Si-SiGe hetero-structures is a concern due to potential intermixing and strain relaxation during the anneals required to activate donors [28]. Ion implantation and activation of phosphorus implants in silicon quantum wells in SiGe structures has been demonstrated [59]. Nuclear spin free Si-SiGe on oxide (SGOI) is a promising platform for implementation of a donor-quantum dot qubit architecture. But donor spin coherence in SiGe might be limited to ~ms by electron-phonon coupling [5]. Coherence limiting factors have to be quantified and compared to values for donors in $^{28}$SOI.

|  | Comments | Refs |
|---|---|---|
| donor $T_{1n} \approx 250\, T_{1e}$ |  | [43] |
| donor $T_{2e} \leq 2\, T_{1e}$ | $T_{2e}$= 10 s in pure $^{28}$Si at 1.8 K | [3] |
| donor $T_{2n} \leq T_{1e}$ | $T_{2n}$=150 s | [42] |
| donor at interface $T_{2e}$=0.3-0.5 ms, $T_{2n} \approx$ tens of ms, ~50 nm from SiO$_2$, 5 K | sensitive to interface quality | [48, 50] |
| quantum dot electron spin $T_{2e}$ | >10 µs | [24, 54] |
| donor electron - nuclear spin state transfer | ~40 ns for entanglement ~10 µs for state transfer | [2, 29, 30] |
| donor – dot exchange coupling | gate controlled with high on/off ratio | [7, 55, 56] |

Table 1: Summary of electron and nuclear spin coherence relationships.

## 4. ELEMENTS OF DEVICE FABRICATION FOR DONOR-DOT SPIN QUBITS

A more readily available, promising substrate for donor-dot device fabrication is $^{28}$SOI, i. e. silicon on insulator where a $^{28}$Si enriched epi layer is grown on a thin natural silicon device layer [38]. Top gates formed by standard e-beam lithography define quantum dots. Careful annealing and re-oxidation steps are required to enhance the oxide quality [60]. Figure 3 shows an electron micrograph of a prototype double quantum dot device with charge sensors. Local accumulation and depletion gates are defined to control dot occupancy with high tuning control.

Single ions can be implanted into double quantum dot devices with scanning probe alignment [61] and single ion impacts can be detected through sensing of changes of the current through a dot or an adjacent charge sensor following single ion strikes [62]. Gate electrodes are formed from metals that can sustain the required post-implantation anneals [38]. Back gate formation requires back etching of the silicon substrate to the box and lithography on the back side. Alignment to features on the top can be achieved with common vias through the device layer and box. Proof-of-concept studies can also be conducted with back gates formed by high energy ion implantation through the device layer and box. While this process does not require the back etching step it might lead to increased noise from damage to the buried oxide and formation of local, sub-micron scale gates is difficult with this process due to the range straggling of MeV ions as they transverse the device and buried oxide layers.

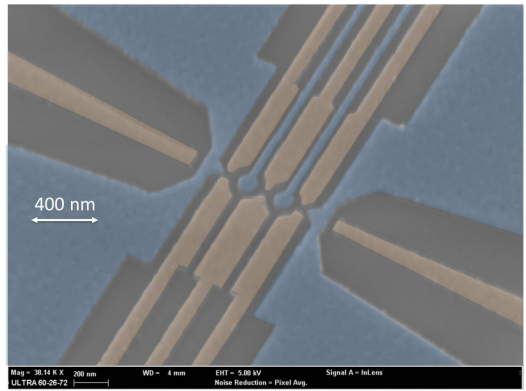

Figure 3: False color electron micrograph of a top gated double quantum dot device with charge sensors and local depletion (brown) and accumulation gates (blue).

## 5. PLACEMENT OF SINGLE DONORS

A great wealth of knowledge has been accumulated on the range and straggling of ions in silicon and the diffusion of donor atoms during thermal annealing [63-65]. For donor spin qubit applications, implant energies have to be selected so that placement uncertainties from range straggling are within tolerances set by the qubit architecture. For the "classical" Kane type quantum computer with nearest neighbor coupling of donors spaced 10 to 20 nm apart, this control of donor placement by ion implantation is still difficult to achieve. Bottom up approaches have sub-nm placement accuracy for single donors [66], but also face challenges, e. g. associated with the bias control range of in plane gates. For ion implantation, the position accuracy is limited by three factors

- Range straggling
- Ion beam spot size
- Diffusion during annealing

Range straggling results from statistical energy loss process during the gradual slow down of ions in the target matrix. Range straggling is reduced for lower ion implantation energies and is reduced for higher projectile mass in a given target matrix. E. g. for implantation of group V donors into silicon, straggling is highest for phosphorus and lowest for bismuth donors. In figure 4 a-c) we show a series of Secondary Ion Mass Spectrometry (SIMS) spectra from low dose implantation of silicon after rapid thermal annealing (RTA). In our example we find that the depth profiles for phosphorus (SRIM depth and FWHM are ~82 nm) and arsenic are dominated by diffusion during RTA. Dopants diffuse through specific mechanisms, e. g. through coupling to vacancies or interstitials. Antimony is a vacancy diffuser, while phosphorous diffuses though coupling to silicon interstitials[65]. The latter are injected form the Si-SiO$_2$ interface during RTA which leads to segregation of P atoms to the interface [67], in a process closely related to oxidation enhanced diffusion [65]. Arsenic couples to both vacancies and interstitials. The observed segregation to the top interface is less severe for arsenic then for phosphorus. Diffusion is minimal for antimony and for bismuth dopants (4 b and c). For antimony, diffusion is retarded by oxidation and makes Sb implants compatible with growth of a high quality oxide following a (single ion) implantation step. For bismuth, which has very low solubility in silicon, dopants can be trapped by re-crystallization of a pre-amorphized silicon layer. This approach requires careful optimization to avoid detrimental effects on donor coherence due to residual damage but it promises to allow activation of bismuth at the required concentrations, with minimal diffusion and thus taking advantage of the smallest straggling of all group V donors in silicon. The FWHM of the Bi donor distribution with a peak depth of 44 nm is just 27 nm. In comparison, the $^{123}$Sb implant in a comparable peak depth of 50 nm has a FWHM of 62 nm, while the 60 keV $^{121}$Sb implant shows a FWHM of 33 nm for a peak depth of 25 nm. These results are in relatively good agreement (especially for trends) but not in full quantitative agreement with standard SRIM simulations [68, 69]. For integration of single donors with quantum dots, range straggling and diffusion have to be controlled so that donors are aligned to single dots and the donor – dot exchange coupling can be controlled for sufficiently fast and precise gate implementation. Requirements for lateral placement are relatively moderate, since typical quantum dots have effective diameters of ~100 nm [26, 28, 60]. Thus a lateral placement uncertainty of even 50 nm allows lateral donor – dot alignment with high success yield. Collimation of the implant ion beam to <50 nm can be achieved with apertures in dynamic shadow masks (see below).

      The required control of donor depth below a quantum dot is more stringent and is given by limits of the tunability of the exchange coupling between donors and an electron in a quantum dot. The characteristic length scale for J coupling is set by the extend of the dot electron wave functions, which is set by the confinement potential and is ~5 to 10 nm in typical top gated quantum dots, similar to the extend of 2DEG's in field effect transistors [70]. For donors in silicon, Bohr radii range from ~1.8 (P, Sb) to 1.5 nm (Bi). J coupling between an electron in a dot and an electron on a donor below the dot can be tuned to have appreciable strengths needed for fast and precise exchange gate execution for distances of ~20 to 40 nm. The requirement for gate execution time is set by the applicability of quantum error correction codes at a given gate fidelity and electron spin coherence time. For dot electron spin coherences times of at least a few tens of µs gates need to be executed within about 1 to 10 ns, requiring J>10 µeV. Both top and back gates can be tuned to displace both the donor and dot electron wave functions in order to turn exchange coupling on and off. The donor depth must be controlled within the J-tuning range and we estimate that ion implantation of donors into a depth of ~25 nm with a FWHM of 20 nm will enable this with high yield. The 60 keV Sb implant from Figure 4 b) shows a FWHM of ~33 nm, and 60 % of donors would be placed in a 30 nm wide depth window from 10 to 40 nm below a quantum dot. Using Bismuth, range straggling is further reduced (Fig. 4 c) [71].

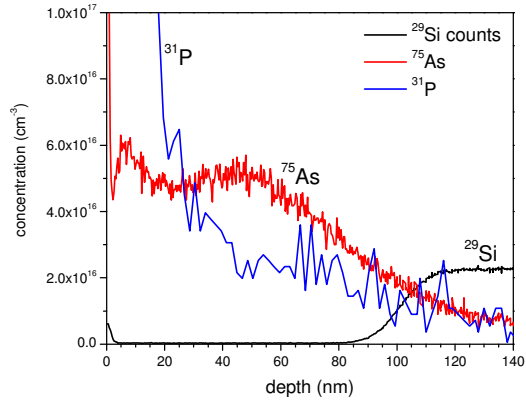
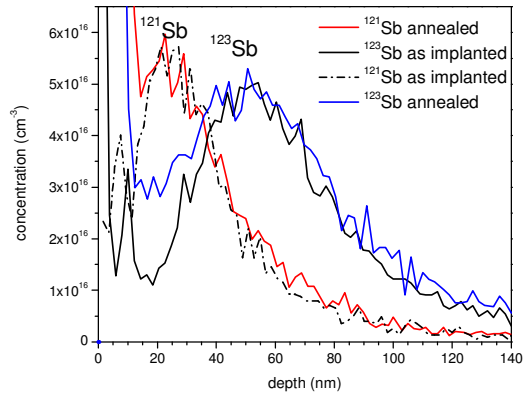
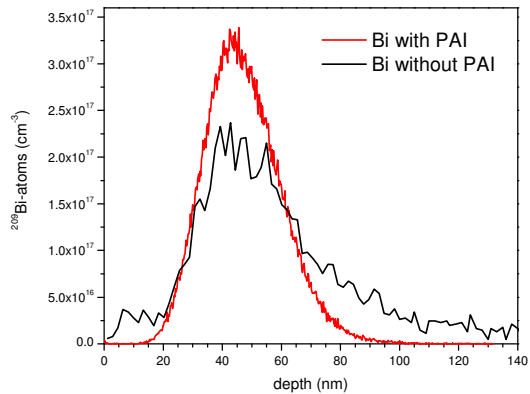

Figure 4 a-c): SIMS depth profiles for P (top, blue, 58 keV, $4\times10^{11}$ cm$^{-2}$, 7 degree, 920 C, 10 s) and As (top, red, 120 keV, $4\times10^{11}$ cm$^{-2}$, 7°, 920 C, 10 s), Sb (middle) and Bi (bottom) implanted into silicon. $^{121}$Sb: 60 keV, $2\times10^{11}$ cm$^{-2}$, $^{123}$Sb, 120 keV, $4\times10$ cm$^{-2}$ (850 C, 10 s) Bi: 120 keV, $10^{12}$ cm$^{-2}$ (with PAI: annealed at 600 C for 4 min., without PAI, annealed at 650 C for 9 min). P and As were implanted into $^{28}$SOI with a 100 nm thick $^{28}$Si enriched epi layer (99.93%). The $^{29}$Si counts in the As profile indicate the transition from the $^{28}$Si enriched to the natural silicon layer.

Donor placement tolerances in a coupled donor – quantum dot architecture are much relaxed compared to requirements for nearest neighbor donor – donor coupling [1] or coupling along a donor chain [72]. Detailed theoretical analysis, analog to the work that has been performed on J coupling between dots and between donors, is required to guide experimental tests and to identify trade-offs e. g. between donor – dot alignment tolerances, coherence times and gate fidelities for donor-dot devices formed in $^{28}$SOI vs. Si-SiGe.

The positions for the placement of each donor atom have to be aligned to pre-formed or to be formed quantum dot electrodes. This can be achieved e. g. through imaging with a scanning force microscope. Here, a small hole in the cantilever near the imaging surface acts as a dynamic shadow mask. Figure 5 shows a schematic of this concept of ion implantation with scanning probe alignment [61, 73, 74]. An example of a dot pattern formed in PMMA by ion implantation (8 keV $Ar^{2+}$) with scanning probe alignment is shown in Figure 6 [75]. Dot diameters are 90 nm and match the diameter of the hole in the cantilever that was used as a dynamic shadow mask.

In ion implantation with scanning probe alignment, a low divergence, broad beam of dopant atoms, e. g. extracted from an electron beam ion trap [61, 76] (if high ion charge states are required to enhance single ion detection), or e. g. from a more common microwave driven ion source [62] reaches the implant chamber after mass selection e. g. in a bending magnet or an ExB filter. The beam with a current density of ~0.1 to 1 nA/mm$^2$ impinges on a pre-collimator in which a micron scale hole is drilled and aligned to a nano-scale hole in the cantilever (Fig. 5). The imaging function of the scanning probe allows non-invasive imaging of the device structure and alignment of the ion beam to regions of interest with a resolution that is limited by the size of the collimating aperture. Using focused ion beam drilling and local thin film deposition, apertures with diameters as small as 5 nm have been formed [77]. Figure 5 also shows an in situ scanning probe image of a silicon nano-wire device. Apertures formed by e-beam lithography above quantum dots can protect against alignment errors in the single ion implantation step.

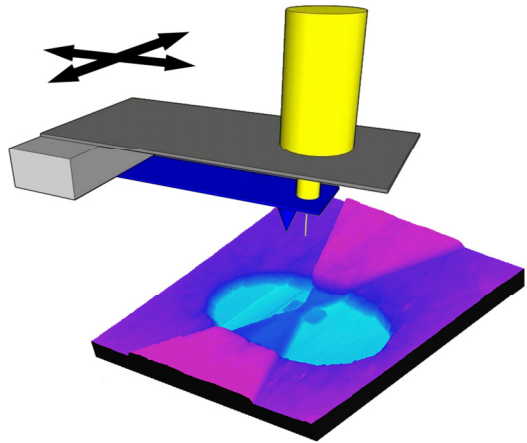

Figure 5: Schematic of ion implantation with scanning probe alignment and an in situ scanning probe image (15x15 μm$^2$) of a silicon nanowire device [73].

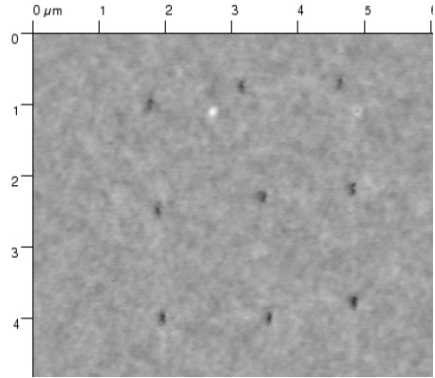

Figure 6: Scanning probe image of a pattern of holes formed by ion implantation with a dynamic shadow mask in PMMA after development. The dot diameter is 90 nm equal to the diameter of the collimator in the cantilever [75].

## 6. SINGLE ION IMPLANTATION

Detection of single ion impacts can be achieved through sensing of upsets in device currents following single ion strikes [62, 73, 74]. Sub-100 nm scale devices are subject to random telegraph noise (RTN) in device currents e. g. due to the random changing of the occupancy of individual defect states at the $SiO_2$-Si interface [38]. RTN effects can often be observed already at room temperature. Given this sensitivity of small devices to individual Coulomb scatterers, it is not surprising that device currents are sensitive to the impact of multi-keV ions, which form many defects when they impinge into the devices. Upon exposure to a test beam of low energy xenon ions, large steps are observed in source drain current of an etched silicon nano-wire device formed on SOI (SOI thickness 50 nm, channel length 300 nm, width 100 nm). These upsets indicate single ion hits (Figure 7). Here, current steps from single ion hits decrease the source-drain current by ~20%. The steps are much increased for these 100 nm scale devices compared to earlier demonstrations of single Sb-ion implantation into transistors with 2x2 $\mu m^2$ channels [62, 74]. Current steps arise when single ions deposit kinetic energy (and potential energy for multiply charged ions) in collisions with target electrons and nuclei as they transverse a thin gate dielectric into the silicon channel. Mostly positively charged defects in the oxide and at the Si-$SiO_2$ interface increase the effective gate voltage in the channel. In earlier studies, we have observed increasing channel currents in transistors with pierced top gates, both in FinFets [73] and planar Fets. For higher implant doses, accumulation of lattice damage degrades mobility. For series of single ion implants, we have observed first a rise in the channel current (from effective gate voltage increases due to positively charged defects) followed by a reversal to stepwise current decreases (due to accumulation of mobility degrading charged defects). Microscopic mechanisms for single ion induced current changes depend on the device geometry, device biasing and implant conditions [62, 74, 78, 79]. The impact of single ions with multi-keV energies is accommodated by the formation of multiple electronic defects which represent a large perturbation in 100 nm scale devices and detection of current changes from single ion hits is a robust method for single ion detection. An alternative method is detection of secondary electrons [61, 80] which has the advantage of not requiring the presence of active electronic device function for the single ion implant step. A disadvantage is that secondary electrons (emitted in a cosine distribution) are in part blocked by any apertures formed on the device layer. The latter can protect against alignment errors and accidental implantation and also aid alignment of the single ion implant step to quantum dots. Damage in the device channels and gate oxides can be annealed following completion of single ion placement. For some donor species, in particular Sb, diffusion is retarded by oxidation, making them also compatible with oxide regrowth.

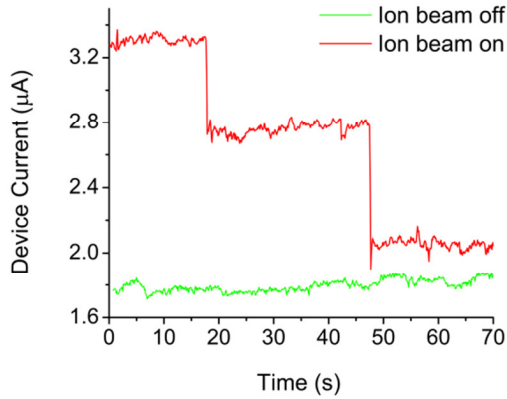

Figure 7: Steps in nano-wire current in response to single ion hits.

## 7. OUTLOOK

We have described a donor-quantum dot architecture that integrates a nuclear spin memory with electron spins in quantum dots. Review of coherence properties of donor electron and nuclear spins in $^{28}$Si highlights the potential for such an architecture for the realization of large scale quantum computers. Critical areas for development include the optimization of gate dielectrics and a better understanding and control of coherence limiting processes for integrated donors and electron spins in quantum dots. We have outlined elements for practical realization of coupled donor – quantum dot devices, including single ion implantation into back gated $^{28}$SOI.

**Acknowledgements**

We thank L. Greenman and K. B. Whaley for fruitful discussions. This work was supported by the US National Security Agency under 100000080295 and by DOE under contract no DE-AC02-05CH11231 (LBNL).